# Voltage controlled core reversal of fixed magnetic skyrmions without a magnetic field

Dhritiman Bhattacharya[1], Md Mamun Al-Rashid[1,2] and Jayasimha Atulasimha[1, 2 *]

[1]*Department of Mechanical and Nuclear Engineering, Virginia Commonwealth University, Richmond, VA 23284, USA*
[2]*Department of Electrical and Computer Engineering, Virginia Commonwealth University, Richmond, VA 23284, USA*
* Corresponding author: jatulasimha@vcu.edu

Using micromagnetic simulations we demonstrate core reversal of a fixed magnetic skyrmion by modulating the perpendicular magnetic anisotropy of a nanomagnet with an electric field. We can switch reversibly between two skyrmion states and two ferromagnetic states, i.e. skyrmion states with the magnetization of the core pointing down/up and periphery pointing up/down, and ferromagnetic states with magnetization pointing up/down, by sequential increase and decrease of the perpendicular magnetic anisotropy. The switching between these states is explained by the fact that the spin texture corresponding to each of these stable states minimizes the sum of the magnetic anisotropy, demagnetization, Dzyaloshinskii-Moriya interaction (DMI) and exchange energies. This could lead to the possibility of energy efficient nanomagnetic memory and logic devices implemented with fixed skyrmions without using a magnetic field and without moving skyrmions with a current.

## I. INTRODUCTION

Ongoing quest for high density and high speed nanomagnetic computing devices has led to the exploration of novel materials, devices and switching strategies. Recently, a topologically protected spiral spin structure called skyrmion has attracted attention due to its potential use as such devices. Skyrmions were first proposed to explain hadrons [1]. Later theories predicted the existence of magnetic skyrmions in the chiral helimagnets [2]. Subsequently, experiments showed the evidence of skyrmion lattices in bulk [3-5] and thin film [6-9]. Here Dzyaloshinskii-Moriya interaction [10-11], given by $H_{DM} = -D_{12} \cdot (S_1 \times S_2)$, that is present in non-centrosymmetric magnets or thin films interfaced with a metal with large spin orbit coupling, stabilizes the skyrmion state. Fig. 1(a) shows a schematic of the magnetic configuration of a skyrmion. Several schemes have been investigated to design racetrack memories [12-14] and logic gates [15] by manipulating the motion of a skyrmion because the pinning current is orders of magnitude less than that of domain walls [16,17]. Although an ultra dense memory device with readout integrated to an MTJ can be assembled via core reversal of stationary skyrmion, such in-situ control of magnetic state of skyrmions has not been studied extensively. Core reversal induced by microwave [18], magnetic field [19], spin current [20], and conversion between Skyrmion and ferromagnetic state have been shown using STM tip [21] and combination of electrical and magnetic fields [22]. However, skyrmion core reversal with voltage controlled magnetic anisotropy has not been shown so far. This paper proposes an extremely energy efficient route for core reversal of a magnetic skyrmion with an electric field.

Previous schemes for switching nanomagnets with perpendicular magnetic anisotropy using VCMA could only achieve a maximum of 90° magnetization rotation (from out of plane to in-plane). In our proposed scheme, core reversal can be achieved without requiring any external field. Furthermore, we can switch reversibly between two skyrmion states and two ferromagnetic states, i.e. skyrmion states with the magnetization of the core pointing down/up and periphery pointing up/down, and ferromagnetic states with magnetization pointing up/down, by sequential increase and decrease of the perpendicular magnetic anisotropy.

Electric field control of magnetization reversal offers an extremely energy efficient route for manipulating magnetisation. VCMA [23-25], strain [26-28], magnetoelectric switching using antiferromagnetic layer coupled to a ferromagnet [29-30] are techniques that offer electric field control of magnetization in nanomagnets. In this paper, we show with rigorous micromagnetic simulation that a skyrmion core can be reversed solely by modulating the perpendicular anisotropy with an electric field, i.e. without any external magnetic field. This utilizes the dependence of the anisotropy value in the ferromagnet/oxide interface (which originates from the overlap between oxygen's $p_z$ and ferromagnet's hybridized $d_{xz}$ orbital [31]) on the electron density [32]. We use this voltage control of magnetic anisotropy (VCMA) technique to create and annihilate skyrmions and as well as to achieve core reversal and switching between ferromagnet and skyrmion states. Such electric field control of topologically stable skyrmions can create a new avenue towards implementing energy efficient nanomagnetic computing.

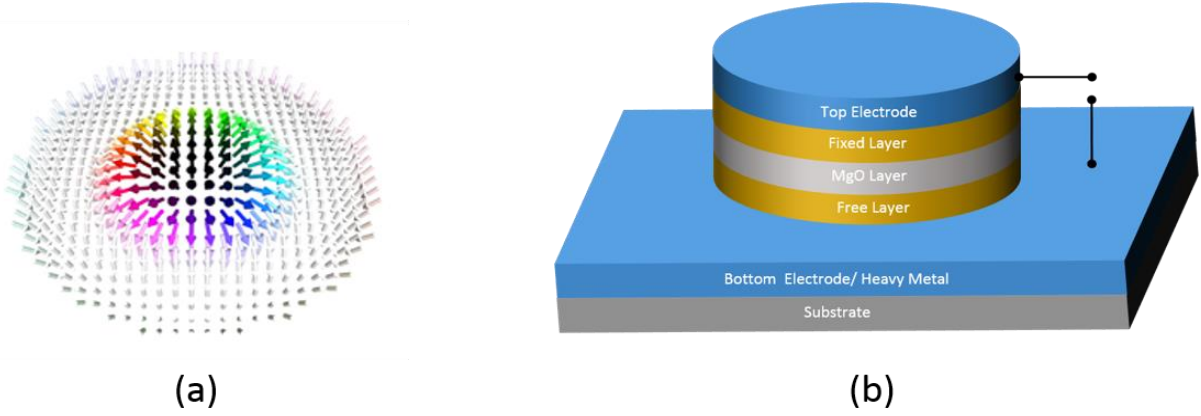

FIG 1. (a) Skyrmion (b) MTJ structure (variants of the readout schemes in the appendix).

## II. SWITCHING SIMULATION

We simulate the magnetization dynamics in a perpendicular anisotropy CoFeB/MgO/CoFeB MTJ structure shown in Fig. 1(b) to demonstrate skyrmion core reversal. The bottom CoFeB layer is the free layer which is chosen to be a nanodisk with radius of 80 nm and thickness of 1 nm. Simulations were performed using the micromagnetic simulation software- Mumax [33]. Our geometry was discretized into 1×1×1 nm$^3$ cells. The change in uniaxial anisotropy constant is realized by modulating the electric field.

In the MuMax framework [33], the magnetization dynamics is simulated using the Landau-Lifshitz-Gilbert (LLG) equation:

$$\frac{\delta \vec{m}}{\delta t} = \vec{\tau} = (\frac{\gamma}{1+\alpha^2}) \times (\vec{m} \times \vec{H_{eff}} + \alpha \times (\vec{m} \times (\vec{m} \times \vec{H_{eff}}))) \quad (1)$$

where $m$ is the reduced magnetization ($M/M_s$), $M_s$ is the saturation magnetization, $\gamma$ is the gyromagnetic ratio and $\alpha$ is the Gilbert damping coefficient. The quantity $H_{eff}$ is the effective magnetic field which is given by,

$$\vec{H_{eff}} = \vec{H_{demag}} + \vec{H_{exchange}} + \vec{H_{anisotropy}} \quad (2)$$

Here, $H_{demag}$, $H_{exchange}$ and $H_{anisotropy}$ are the effective field due to demagnetization energy, the effective field due to Heisenberg exchange coupling and DMI interaction, and the effective field due to the perpendicular anisotropy evaluated in the MuMax framework in the manner described in Ref [33].

Typical parameters for the CoFeB layer are listed in table 1. [14, 34]

TABLE 1.

| Parameters | Value |
| --- | --- |
| Saturation Magnetization ($M_s$) | $1 \times 10^6$ A/m |
| Exchange Constant (A) | $2 \times 10^{-11}$ J/m |
| Perpendicular Anisotropy Constant ($K_u$) | $8 \times 10^5$ J/m$^3$ |
| Gilbert Damping ($\alpha$) | 0.03 |
| DMI Parameter (D) | 3 mJ/m$^2$ |

## III. DISCUSSION

The reversal of the skyrmionic state is achieved through modulation of the perperdicular magnetic anisotrpy (PMA) by applying an electrical volatge. Modulation of the PMA initiates a change in the orietation of the spins and ultimately the equilibrium spin configuration is determined by *minimizing* the total energy of the system which includes exchange energy, DMI energy, magnetic anisotropy energy and demagnetisation energy. *We note that the micromagnetic simulation describes the evolution of the magnetic configuration with time to reach this local minimum.* The reversal is a two step process. The voltage profile and anisotropy energy density change with time, the magnetic energies of the system at various states and configurations of different magnetic states visited during the switching process is shown in fig 2 (a), fig 2 (b) and fig 2 (c) respectively.

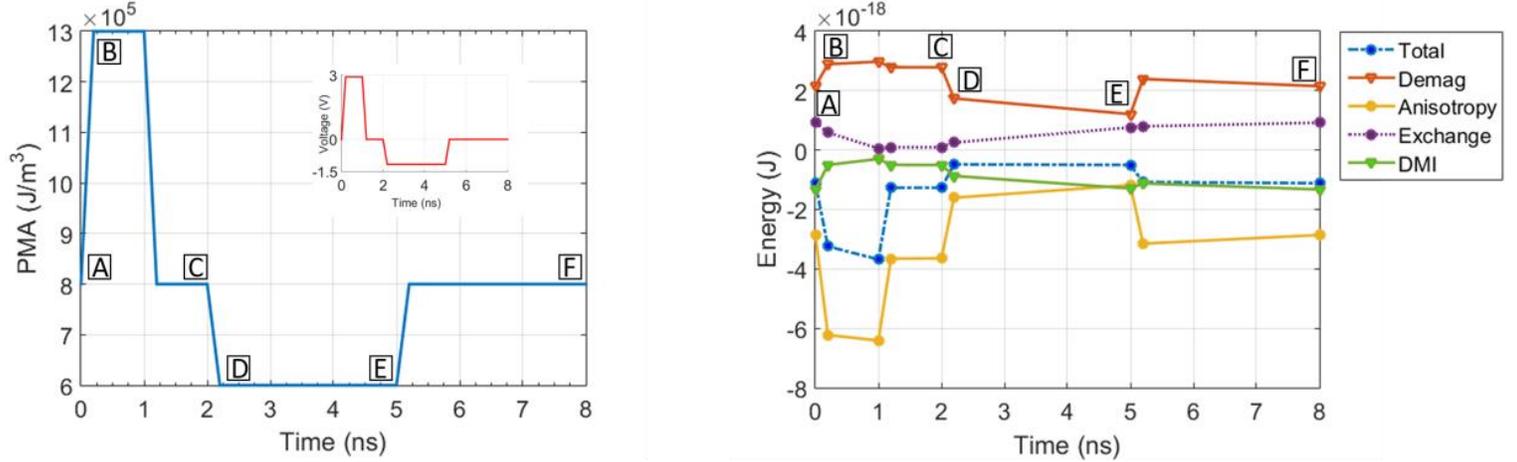

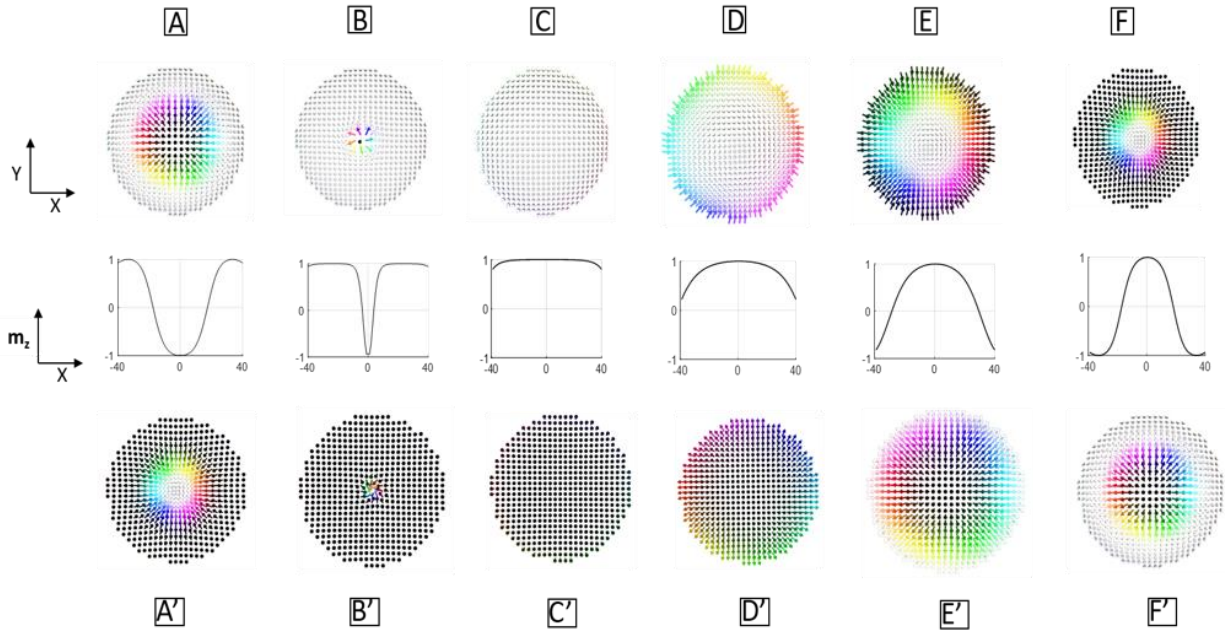

FIG 2. (a) Anisotropy energy density and voltage vs time, (b) Energy vs time, (c) spin states at different time and associated magnetization component in the z-direction of different points along the diameter.

Let us assume we start with a skyrmion with core pointing down (Fig 2 (c), state A). In the first step, a positive voltage is applied to the skyrmion which strengthens the perpendicular anisotropy. This forces more spins to point in the direction perpendicular to the x-y plane (i.e. in the direction ±z) to reduce the anisotropy energy. Majority of the spins are pointing up (+z) in the initial A state because the peripheral region with spins predominantly up is roughly three times the area of the core with the spins predominantly pointing down. Thus, the exchange interaction makes the +z direction the preferred direction among the two possible perpendicular spin orientations (±z). This results in reduction of the skyrmion diameter (fig 2(c), state B) with increasing PMA. With strong enough perpendicular anisotropy, the skyrmion state can no longer persist and all spins point in the +z direction forming a complete ferromagnetic state. Increase in the DMI and demagnetization energy due to this transformation is compensated by the reduction in anisotropy and exchange energy as shown in fig 2 (b). This ferromagnetic state is also stable (similar to the skyrmionic state A) and remains in this state when the applied voltage is reduced to zero (fig 2(c), state C). This is what makes it non-volatile.

Next, the perpendicular anisotropy is lowered by applying a negative voltage. With sufficiently large negative electric field, the DMI and demagnetization energies prevail over the anisotropy and exchange energy. This causes the spins to rearrange themselves in order to minimize the total energy resulting in the transformation from a state where almost all the spins were pointing up (fig 2(c), state C), to a state where the spins orient to create an incomplete skyrmion (skyrmion number between 0.5 and 1) with the core pointing up and spins in the perimeter of the disk tilted in the –z direction (fig 2(c), state E). The tilting starts at the periphery of the disk because this results in a smaller penalty in terms of exchange energy than the tilting of the spins in the core. If the perpendicular anisotropy is restored by removing the applied electric field, spins in the perimeter completely rotates to the –z direction, while the core spins still point up. Thus, a skyrmion with core pointing up, periphery with spins pointing down is formed, which is the new stable state (fig 2(c), state F). This is also non-volatile. Hence, we have bistable skyrmionic state "0" and "1". *We note that each equilibrium configuration (A, C, F) was attained by forming a magnetic configuration that corresponds to a local energy minimum closest to its prior state, i.e. the state from which this system evolves, and separated from other local minima by an energy barrier.* Thus, when the system evolves from a state stabilized by high PMA due to VCMA with a positive voltage, it settles to the ferromagnetic state when the VCMA is reduced to zero. But, when the system evolves from a state stabilized by low PMA due to VCMA with a negative voltage, it settles to the skyrmion state when the VCMA is reduced to zero. But it cannot spontaneously switch between the skyrmion and ferromagnetic state due to the energy barrier separating them.

Had we started with a skyrmion with core pointing up and applied the same voltage pulse, we would end up with a skyrmion with core pointing down. This states (A'-F') are shown in fig 2(b). Also, if the initial state is ferromagnetic, we can apply a negative pulse to stabilize a skyrmion. Thus we can reversibly switch between these four states (two skyrmionic and two ferromagnetic states). This behavior is explained in Fig. 3. Being able to toggle between the two stable skyrmionic states or ferromagnetic states make it a viable memory element.

## IV. CONCLUSION

In summary, our simulations have demonstrated the use of voltage controlled magnetic anisotropy for core reversal of a magnetic skyrmion, skyrmion mediated ferromagnetic state reversal and switching between skyrmion and ferromagnet states without requiring any bias magnetic field. More involved simulation that include thermal noise could establish the room temperature stability of various states and error involved in switching between these states, but is beyond the scope of this paper. Possible methods to read the different states by integrating a magnetic tunnel junction (MTJ) can be achieved in the manner of Fig 1 (b) of this paper as well as Fig A 1 (a) and (b) discussed in the appendix.

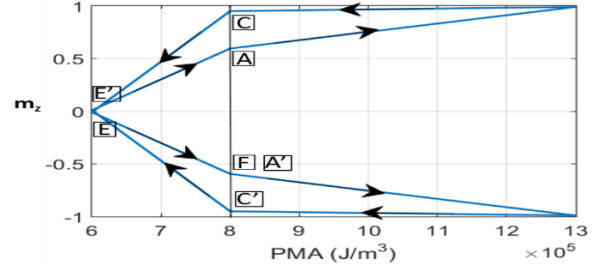

FIG 3. Normalized perpendicular magnetization ($m_z$ vs. PMA). Arrows show the transition directions between the states.

We can estimate the energy dissipated in switching between the skyrmions states as follows: the modulation of the interface anisotropy energy is given by $J_{sa}=J_o + a\,E$, where $a$, E and $J_0$ are respectively the coefficient of electric field control of magnetic anisotropy, the applied electric field and the interface anisotropy energy at zero bias field. Reported values of coefficient of anisotropy energy change ($a$) by electric field are ~11-33 uJ/m$^2$ per V/nm [25]. Insertion of Hafnium (Hf) monolayer can increase it by ~5.2 times [35]. Thus, with a 1 nm thick free layer, $1.71 \times 10^5$ J/m$^3$ change in the anisotropy energy density can be obtained per volt. Our desired anisotropy change can be achieved in the range of -1.17 V to 2.92 V. This values translates into an energy dissipation of ~1.8 fJ per switching cycle if all the energy required to charge the capacitive MgO layer (relative permittivity ~7 [36], thickness ~1nm, diameter ~ 80 nm) is ultimately dissipated. Further scaling of the free layer (skyrmion) and therefore the oxide layer to ~40 nm diameter could decrease the energy dissipation to ~ 450 aJ that would be comparable to conventional CMOS devices [37]. However, an advantage of the nanomagnetic element is its non-volatility. Further, substantial reduction of energy dissipation may be achieved by lowering the electric field needed for the switching process if the coefficient of anisotropy energy change ($a$) is enhanced in future materials/interfaces. Moreover, we can switch between states in a few nanoseconds, which is competitive for computing applications, particularly given low energy dissipation and non-volatility.

These theoretical results could stimulate experimental work on switching fixed skyrmions with voltage controlled magnetic anisotropy, novel interfaces with higher $a$ and lead to implementation of energy efficient memory devices based on skyrmion core reversal or ferromagnetic state reversal (via an intermediate skyrmion state), and Boolean and non-Boolean magnetic tunnel junction (MTJ) based computing in the manner of Ref [38] and [39] respectively.

**ACKNOWLEDGEMENTS**

D.B, M.M.A. and J.A are supported in part by the National Science Foundation CAREER grant CCF-1253370.

**APPENDIX A**

Read out schemes: The readout scheme proposed in the main paper (shown in Fig 1 (b)) enables direct readout of the skyrmionic states. While this is ideally suited to distinguish between the two ferromagnetic states as well as the two the skyrmion state, the magnetoresistance ratio between the two skyrmion states is low as the magnetization in the core and the periphery are anti-parallel. Improvement of this "on-off" ratio for just reading the skyrmion states (while also maintaining a high on-off ratio for the ferromagnetic states) can be achieved by using methods to read either the core or the periphery. Fig 4 shows a 3-terminal device with concentric electrodes which enables easy readout of a skyrmion core. The top electrode (writing electrode for VCMA) is a cylindrical shell with inner diameter of 20 nm and outer diameter of 80 nm. Thus, the VCMA affects the region only underneath this electrode. The fixed layer has a diameter of 15 nm and is located above the core. Such a small MTJ hard layer can be fabricated without losing thermal stability [40]. The readout electrode is also 15 nm in diameter and is delineated on top of the fixed layer. Another alternative is to read the magnetization state of the periphery. Fig 5 shows a device structure which fulfills this purpose. This is a two terminal device where one electrode is used for both reading and writing. The read/write electrode and the fixed layer are both cylindrical shells (one above the other) with inner diameter of 20 nm and outer diameter of 80 nm. The physics behind the reversal by applying electric field only in the peripheral region does not differ from the one described in the main paper. Different magnetic states visited in response to perpendicular energy density change with voltage (Fig 6) applied in the peripheral region only are shown in Fig. 7. Note that we used a slightly different voltage pulse in the latter two methods to switch the skyrmion.

Other ways to control magnetic anisotropy: In addition to directed voltage control of magnetic anisotropy one could modulate the magnetic anisotropy of magnetostrictive nanomagnets with acoustic waves or voltage generated strain in a piezoelectric layer in elastic contact with the nanomagnet. However, assuming optimistically that ~100 MPa could be generated in a Terfenol-D nanomagnet with magnetostriciton ~ 1000 ppm[41] the effective change in magnetic anisotropy energy density ~$10^5$ J/$m^3$ can be achieved. This is about one order of magnitude smaller than the required change in anisotropy. Hence, strain mediated switching of skyrmion states could be feasible only if materials that have much larger magnetoelastic coupling and sufficient DMI (when interfaced with appropriate materials] to form skyrmions are developed.

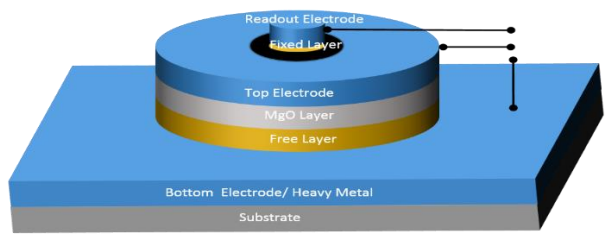

FIG 4. MTJ structure to readout the skyrmion core only

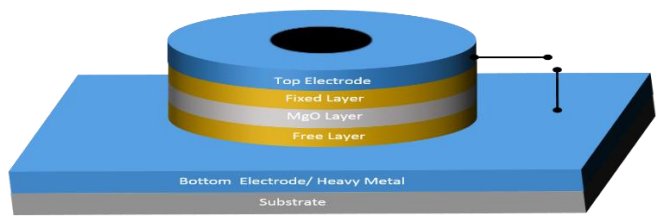

FIG 5. MTJ structure to readout the skyrmion periphery only

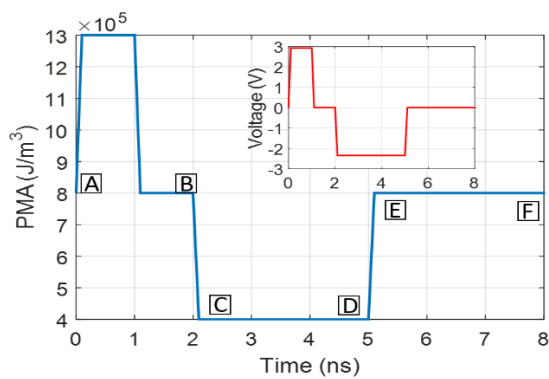

FIG 6. Anisotropy energy density and voltage vs. time for switching using the outer ring only.

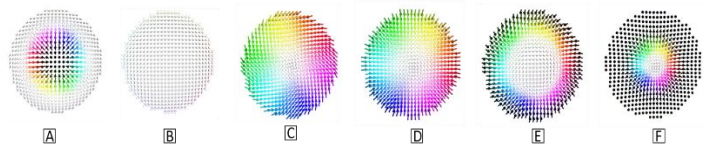

FIG 7. Spin states visited at different times during core reversal using outer ring only.